\documentclass[11pt]{evn}
\usepackage[T2A]{fontenc}
\usepackage[cp1251]{inputenc}
\usepackage{epsf}
\usepackage[russian,english]{babel}
\usepackage{graphicx}
\long\def\maintitle#1{{\vskip 20mm \begin{center}\section*{#1}\end{center}\nopagebreak[4]}}

\long\def\author#1{{\begin{center}\normalsize{\bf#1}\end{center}\vskip-1em\index{#1}}\nopagebreak[4]}
\long\def\address#1{{\begin{center}\small\noindent#1\end{center}\vskip-8mm}\nopagebreak[4]}

\begin{document}
\noindent\mbox{\small The 13$^{th}$ EVN Symposium \& Users Meeting Proceedings, 2016}

\maintitle{TRANSIENT SCIENCE WITH THE e-EVN}

\author{Zsolt~Paragi$^{1}$}

\address{$^{1}$Joint Institute for VLBI ERIC (JIVE), Dwingeloo, The Netherlands}

\begin{abstract} I briefly review transient research with the EVN, with particular emphasis on 
the science that was (or is being) made possible with the latest real-time e-VLBI developments.

\end{abstract}

\section{Introduction}

The European VLBI Network commenced real-time electronic VLBI operations
through a limited number of dedicated 24-hour observing sessions with a subset of the
telescopes in 2006; this array is sometimes referred to as the e-EVN \cite{1}.
The technical advances were accompanied by new policies that allowed
for an easier access to the e-EVN for an increasing number of Target of Opportunity 
(ToO) observations. A well-developed calibration pipeline \cite{2} ensures
rapid access to the initial results, and the implementation of the 
EVN Software Correlator at JIVE (SFXC) allows for a broader range of science topics \cite{3}.
These developments guarantee flexible access to the most sensitive standalone VLBI
network, making a wide range of transient science possible. The first rapid report of 
science observations (The Astronomer's Telegram) \cite{4} and the first refereed journal
papers \cite{5,6} appeared already in 2007, and these were followed by many others.
It is however important to stress that it is not simply real-time correlation that makes the e-EVN
a unique facility; it is the easy and open access, the high-level of support, and especially 
the greatly increased flexibility to follow-up transient phenomena (decreasing response time),
as well as the ability to take part in multi-band campaigns outside of the regular 
EVN observing sessions.

\section{The radio transients parameter space}

In order to discuss the relevant science cases with the broader astronomical community,
we organized a Lorentz Center workshop \lq\lq Locating Astrophysical Transients'' between 
13--17 May 2013, in Leiden, the Netherlands. During this workshop the revolutionary change
in the study of optical transients was set as an example. Before the advent of the major optical
surveys like the Palomar Transient Factory, most of the known optical transients fitted into
either the explosive (supernovae) or eruptive  (novae, luminous blue variables) categories 
\cite{7}.  The large field-of-view, sensitive surveys that probe a range of timescales have
revealed new types of transients. 
One may expect to see a similar revolution in the radio regime in the era of Westerbork/APERTIF 
and the Square Kilometre Array (SKA), to name the two most relevant potential source of 
radio triggers for the e-EVN. The mid-frequency component of the first-phase SKA 
(SKA1-MID) will also be part of global VLBI networks as a phased-array element \cite{8}. 
 
\begin{figure}[t]
\begin{center}
 \vspace*{-2.0 cm}
 \hspace*{-1.3 cm}
\includegraphics[angle=-90,width=6.0in]{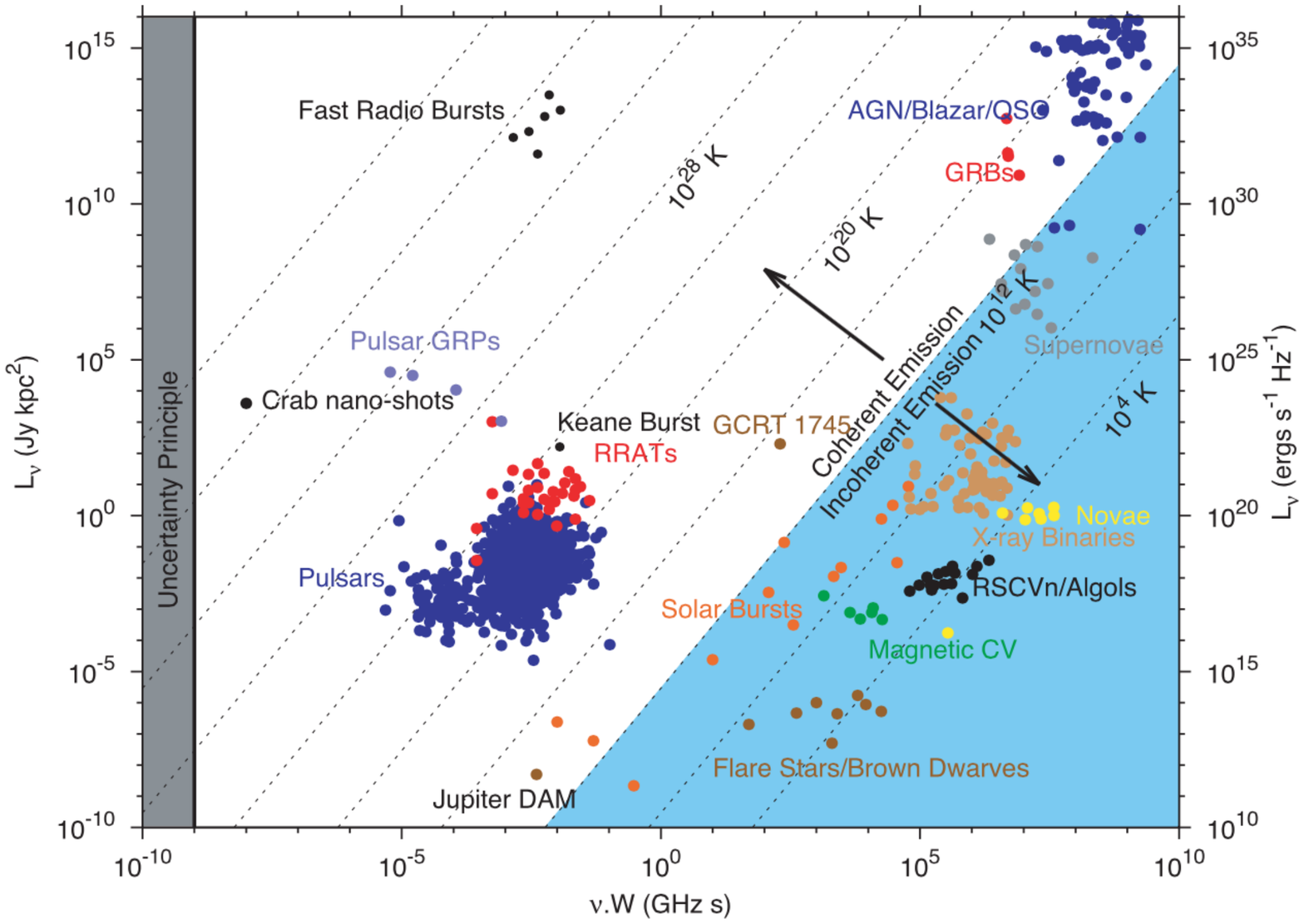} 
\vspace*{-1.0 cm}
\caption{Known types of radio transients shown in the phase space of specific luminosity versus product of 
observing frequency and transient duration (\cite{9}, Fig.~5). Note the division between the 
\lq\lq slow'' synchrotron transients, including typical EVN targets, and the \lq\lq short'' transients producing 
incoherent emission, little studied with VLBI so far. The blazar and GRB flare events are not corrected
for Doppler-boosting.}
\label{fig2}
\end{center}
\end{figure}

The radio transients parameter space is shown in Figure~\ref{fig2}. The events with a duration of more
than a few seconds, up to several years are referred to as slow transients. These have intrinsic brightness 
temperatures below the inverse Compton limit ($T_{\rm b}<10^{12}$~K), and they are mostly due to 
incoherent synchrotron emission. Among these sources we find several targets of VLBI interest. 
The events with a duration of much shorter than a few seconds are often referred to as short transients,
and their emission originates in coherent processes. The Galactic counterparts of millisecond timescale 
pulses are neutron stars, like pulsars and Rotating RAdio Transients (RRAT). Fast Radio Bursts (FRB) 
show dispersion measures  (DM) well above Galacitc values \cite{10}, but their positions have been too 
poorly constrained to date to prove their extragalactic origin.

One of the Leiden workshop conclusions was that increasing the available observing time for real-time 
e-VLBI and shortening the e-EVN reaction times to triggers (on the long term, making automated triggering 
possible) are the ways forward to be able to follow-up a broader range of slow-transient types (Sect.~3).
It was also realised that the FRB field evolves rapidly, and sub-arcsecond localisation of these events is a 
key to understand the nature of these sources. This requires a completely different observing approach and
the need to adopt dispersed single-pulse search techniques for the e-EVN (Sect~4).

\section{Synchrotron (slow) transients }

\subsection{Explosive extragalactic phenomena}

The obvious advantage of VLBI is source localisation at unprecedented precision. While for the 
astronomical interpretation arcsecond localisation would be sufficient in most cases, VLBI data have 
the great potential to distinguish between flaring AGN activity or other types of near-nuclear transients.
The highest resolution ground-based VLBI measurements can probe non-thermal emission brightness 
temperatures up to about $10^{12}$~K, measure tiny displacements (in the 10--100 $\mu$as regime) 
due to a source structural change or proper motion, and can be very helpful to measure compact 
source total flux densities in fields with strong as-scale diffuse emission in the host galaxy. This allows
the study of a broad range of astrophysical phenomena in the Local Universe ($d \leq 200$~Mpc, or 
$z\sim0.05$). For 1~mas resolution the corresponding linear size is $\sim1$~pc at $z=0.05$, 
therefore sub-pc structures can be probed. Within this distance beamed relativistic ejecta can be 
resolved within a few months time, and even mildly relativistic phenomena can be studied on similar
timescales up to a few tens of Mpc distance. Here we assume S/N$\gg$10, resulting in down to 
$\simeq 0.1$~mas reliable size measurement (cf. \cite{11}), and $\leq 0.1$~mas ejecta localisation 
(see e.g. \cite{12}). Below we give a few examples of past (pre-dating e-VLBI) and recent VLBI results
on various types of slow transients.

\begin{figure}[t]
\begin{center}
\vspace{-2.0 cm}
\includegraphics[angle=-90,width=5.0in]{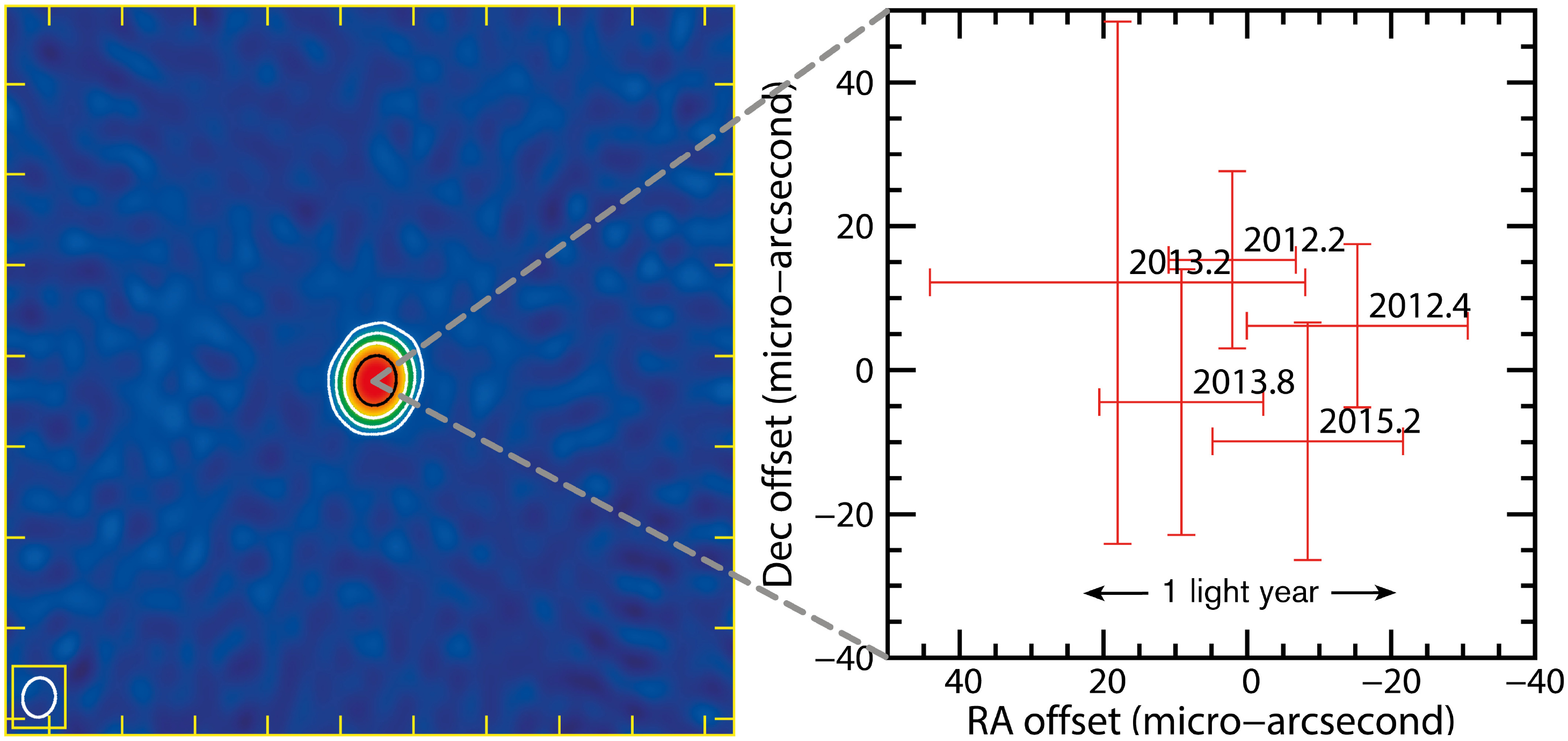} 
\vspace{-1.0cm}
\caption{EVN precision astrometry shows no proper motion in the TDE Swift~J1644+5734 over three years
period of time, constraining the average apparent velocity  $\beta_{\rm app}<0.27c$ 
(\cite{12}, from parts of Fig.~1 and Fig.~2).}
\label{fig3}
\end{center}
\end{figure}

{\bf Supernovae (SNe) and gamma-ray bursts (GRB)} are primary targets for VLBI because their 
resolved radio emission allows for probing the physical properties of their ejecta as well as their 
surrounding environment. Decelerating shells of ejecta have been resolved in EVN, VLBA (and 
global-VLBI) observations for the nearby Type~2 core-collapse events SN\,1979C \cite{13}, 
SN\,1986J \cite{14} and SN\,1993J \cite{15,16}. Nearby starburst galaxies provide a great opportunity
to study a a number of these events in the radio simultaneously, unaffected by dust obscuration.
The nuclear region of Arp220, Arp299A and M82 for example reveal a rich cluster of supernova remnants
and young radio supernovae, some with peculiar properties \cite{17,18,19,20}. Asymmetric, mildly-relativistic 
ejecta have been suggested for some core-collapse types but these have not been confirmed \cite{21,22}. 
Type~Ib/c SN\,2008D has also been initially suspected to have a mildly relativistic jet, but it faded quickly before 
it could be resolved with the EVN \cite{23}. It appears that --most likely-- none of these latter candidates were 
in fact \lq\lq engine driven'', i.e. powered by a black hole formed in the collapse of a massive star, analogous 
to the highly-relativistic GRBs. Only about 1\% of all Type~Ib/c supernovae belong in this category, a
notable example being the broad-line Type~Ib/c SN\,2009bb \cite{24}. While we have not resolved 
any of the GRB afterglows with VLBI besides GRB~030329 \cite{25} yet, e-EVN data in multi-band campaigns
provide useful constraints on the shock physics and the properties of the environment into which it is 
expanding (e.g. \cite{26}).

{\bf Tidal Disruption Events (TDE)}, where tidal forces of a supermassive black hole (SMBH) disrupt 
a star, have received a lot of attention in the past few years. In this process dormant black holes are activated 
to detectable levels as they accrete part of the gas that formed the star. This allows us to study the low-mass 
end of the SMBH population down to $10^6 M_{\odot}$ and possibly below. TDE have initially been detected 
in X-ray surveys \cite{27}, with no trace of radio emission. But the recent detection of a prominent 
flare in Swift~J1644+5734 from $\gamma$ rays to the radio indicated that a fraction of TDE may form 
relativistic radio jets. What this fraction is and how relativistic these jets are is important to know
in order to estimate the expected number of detectable TDE with the SKA, as SKA surveys
may provide an unobscured view to these events \cite{28}. Ultra-high precision astrometry with the
EVN revealed that over three years there was no detectable proper motion in the radio ejecta of
Swift~J1644+5734, providing strong constraints on the intrinsic jet speed as a function of the 
viewing angle \cite{12} (see Fig.~\ref{fig3}). However, another EVN result in ASASSN-14li indicated 
resolved ejecta, although in that case the AGN had already been active before the TDE occured, 
and it is not clear at the moment whether we see a newly formed jet or not \cite{29}. 

\subsection{Galactic transients}

{\bf Black hole X-ray binaries (XRB)}~-- sometimes called \lq\lq microquasars''~-- have been primary
EVN targets since the earliest days of its existence, and they were among the first objects studied by
real-time e-VLBI \cite{5,6}. But in the past it took considerable effort to reveal rapid structural changes 
in XRB: the most prominent steady radio-jet source SS433 was monitored for 6 days in 1985 and 1987, 
and this required practically all available EVN magnetic tapes \cite{30}. The 1987 monitoring produced a 
spectacular show of a series of ejecta during a bright flaring event \cite{31} -- this was just pure luck. 
Real-time e-VLBI does not depend on recording medium at the telescopes, therefore target of opportunity 
or triggered (a proposal activated when certain trigger conditions are met for known transients, e.g. radio 
flux density and/or changes in X-ray state) observations can be scheduled in a much more flexible way 
today. Availability of telescope time can still be a limitation when using the e-EVN, and some of the 
monitoring projects are carried out jointly with the Very Long Baseline Array (VLBA; e.g. \cite{32,33}). 
These project have also demonstrated the power of e-VLBI to quickly readjust the observing schedules 
e.g. when suitable calibrators are found closer to the target (see e.g. Sect.~2 in \cite{33}).

\begin{figure}[t]
\includegraphics[angle=0,width=5.0in]{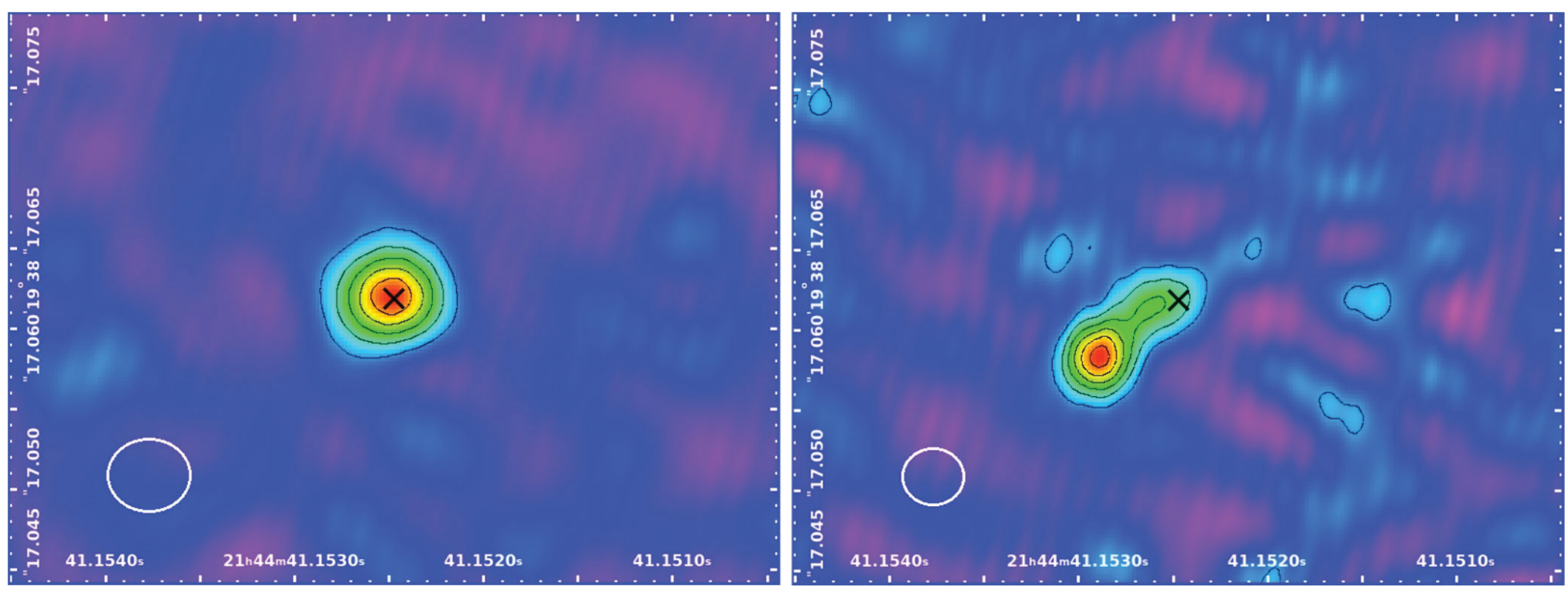} 
\caption{5 GHz e-EVN images separated by one day show structural changes in the low-mass XRB Cyg~X-2,
following an X-ray flare (\cite{35}, Fig.~1).}
\label{fig4}
\end{figure}

In the past fifteen years we have learned a lot about stellar black hole systems. And while it is still not clear
just how relativistic black hole XRB may be, the strong connection in the accretion/ejection physics between 
stellar mass and supermassive black holes has been demonstrated by the discovery of the fundamental 
plane of black hole activity \cite{34}. To understand accretion processes better, our attention now turns toward
the lowest activity states, and also to low-mass XRB systems in which the compact object is a neutron star. 
For example, e-EVN observations have shown radio ejecta in Cyg~X-2 following an X-ray flare 
(Fig.~\ref{fig4}; \cite{35}). But compact jets may also appear in transitional millisecond pulsars when they 
enter the accreting state --~similar to that in low-mass XRB~-- evidenced by the flat spectrum and variable
radio emission in PSR J1023+0038 \cite{36}.

{\bf Stellar-mass isolated black holes (IBH)} may be numerous ($>10^8$) in the Galaxy, and these may get
active during short-period accretion events  \cite{37}. The Solar neighbourhood would contain $\sim$35000 IBH 
in a radius of 250 pc if this population exists. While the predicted X-ray and radio emissions are below the 
current survey limits ($10^{11}$~erg~s$^{-1}$ and 1 mJy, respectively), these sources would show up in 
deep radio observations via their very high proper motion. The brightest of these (few tens of $\mu$Jy) could
be identified with the EVN through their observable proper motion within a day. This will require either systematic
searches in large-field of view EVN data, or triggering from next generation X-ray missions.

\begin{figure}[t]
\vspace{-1.0 cm}
\includegraphics[angle=0,width=5.0in]{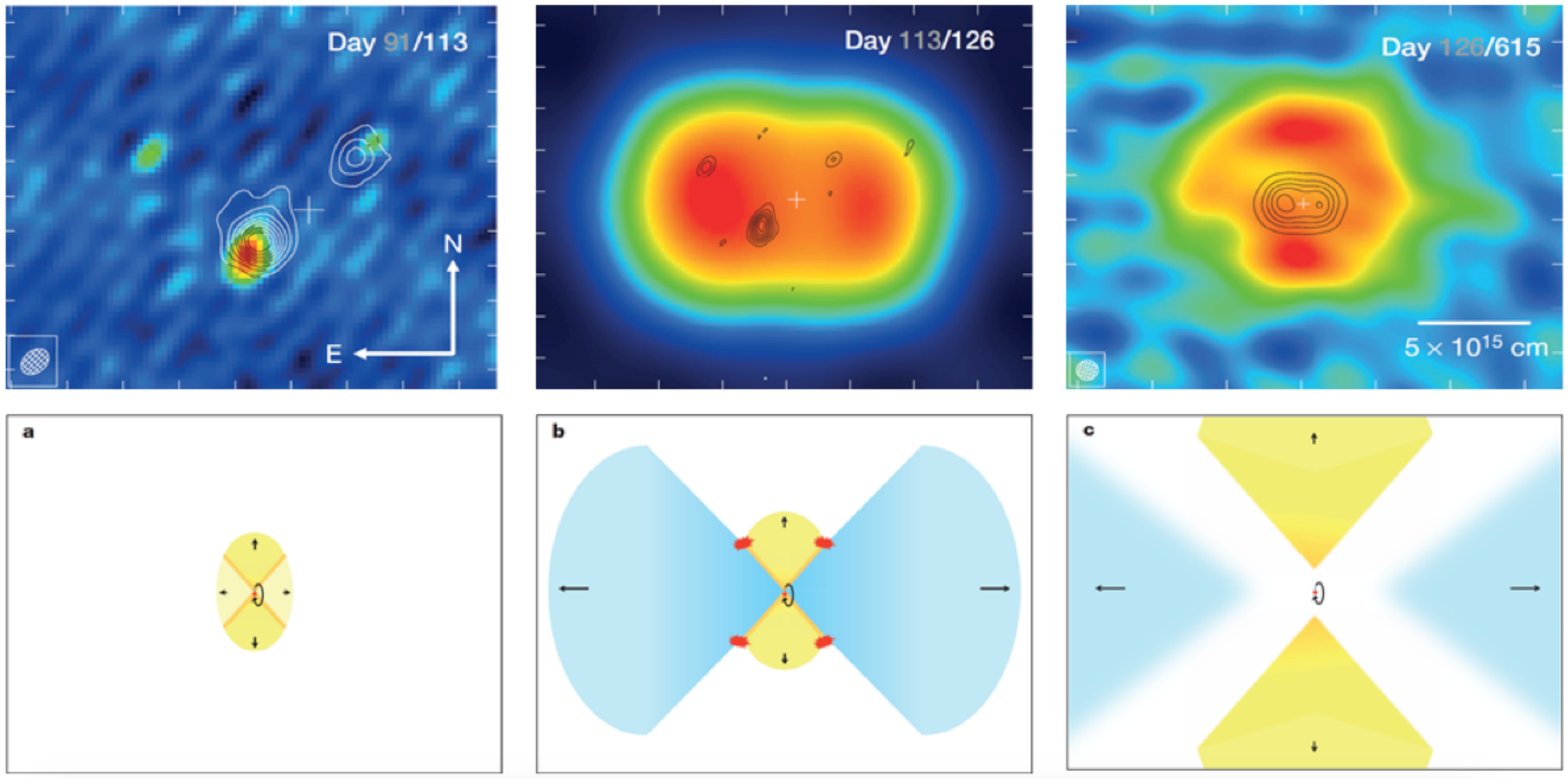} 
\vspace{-1.5cm}
\caption{Various phases of V959 Mon ejecta evolution. Above: comparison of e-EVN (day 91/113)
and VLA (day 126/615) images. Below: schematic illustration of the dense, slow outflow in the orbital plane, 
and the fast polar outflow; their interaction causes the shocks seen on VLBI scales (\cite{42}, re-ordered 
Fig.~2 and Fig.~3).}
\label{fig5}
\end{figure}

{\bf Cataclysmic variables} are binary stellar systems of an accreting white dwarf and a donor star. These
produce prominent UV and X-ray flares in thermonuclear runaway events, but classical novae are strong
radio emitters too. A great example is the recurrent nova RS~Oph. Early EVN observations during
its 1985 outburst revealed non-spherical  synchrotron ejecta \cite{38,39}, a rather controversial result at the
time, because the physical model of novae assumed spherical explosions. This was later confirmed with
the EVN and MERLIN during the 2006 outburst of RS~Oph, indicating a bi-polar, collimated outflow \cite{40}. 
Further support to non-spherical ejecta in novae came with the discovery of V407~Cyg in 2010 \cite{41} and 
V959~Mon in 2012 \cite{42} that, quite unexpectedly, produced $\gamma$-ray flares as well. Monitoring observations
of V959~Mon on a wide range of angular scales with the VLA, MERLIN and the e-EVN have shown a complex 
structure that was interpreted as the result of a slow outflow in the orbital plane and a fast polar wind; 
the interaction of these two regions would cause shocks (detected on milliarcsecond scales with the e-EVN) 
which could be the source of  the $\gamma$ rays as well (see Fig.~\ref{fig5}; \cite{42}). Nowadays it still takes a 
considerable effort to arrange for multiple radio facilities to monitor similar phenomena on a range of angular 
scales. In the near future, a merged e-EVN/e-MERLIN array and, ultimately, SKA-VLBI  \cite{8} will do a 
fantastic job as a single array. 
Besides classical novae, the great e-EVN sensitivity allows us to detect transient radio emission from dwarf novae
as well; the flexible access (rapid response to triggers) is particularly relevant here, because the radio flares 
are short-lived (maximum 1--2 days). A precise parallax distance measurement of SS~Cyg with the VLBA and 
the e-EVN resolved a long-standing discrepancy between accretion theory and previous observations \cite{43}.

\begin{figure}[!ht]
\vspace{-1.0 cm}
\includegraphics[angle=0,width=5.0in]{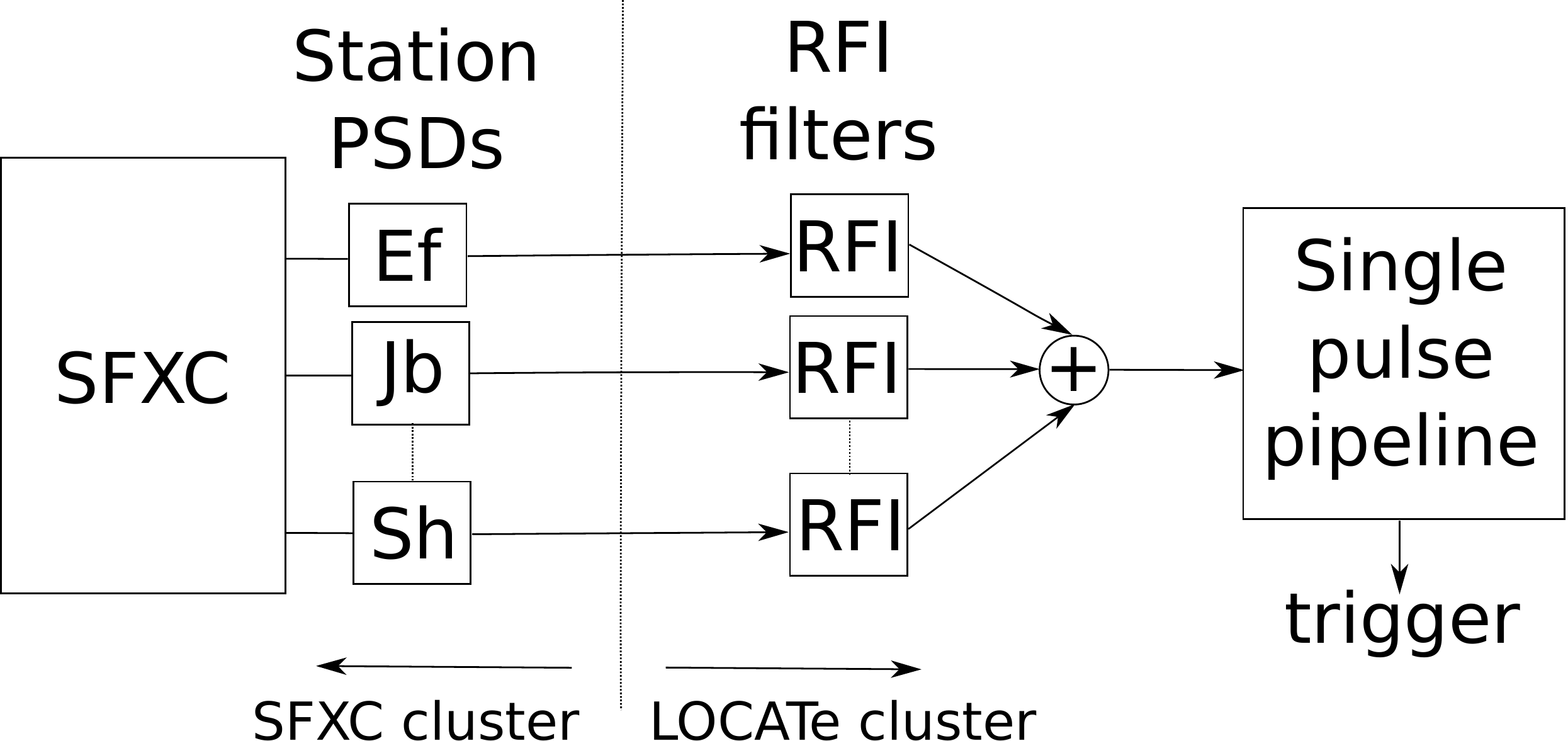} 
\includegraphics[angle=0,width=5.0in]{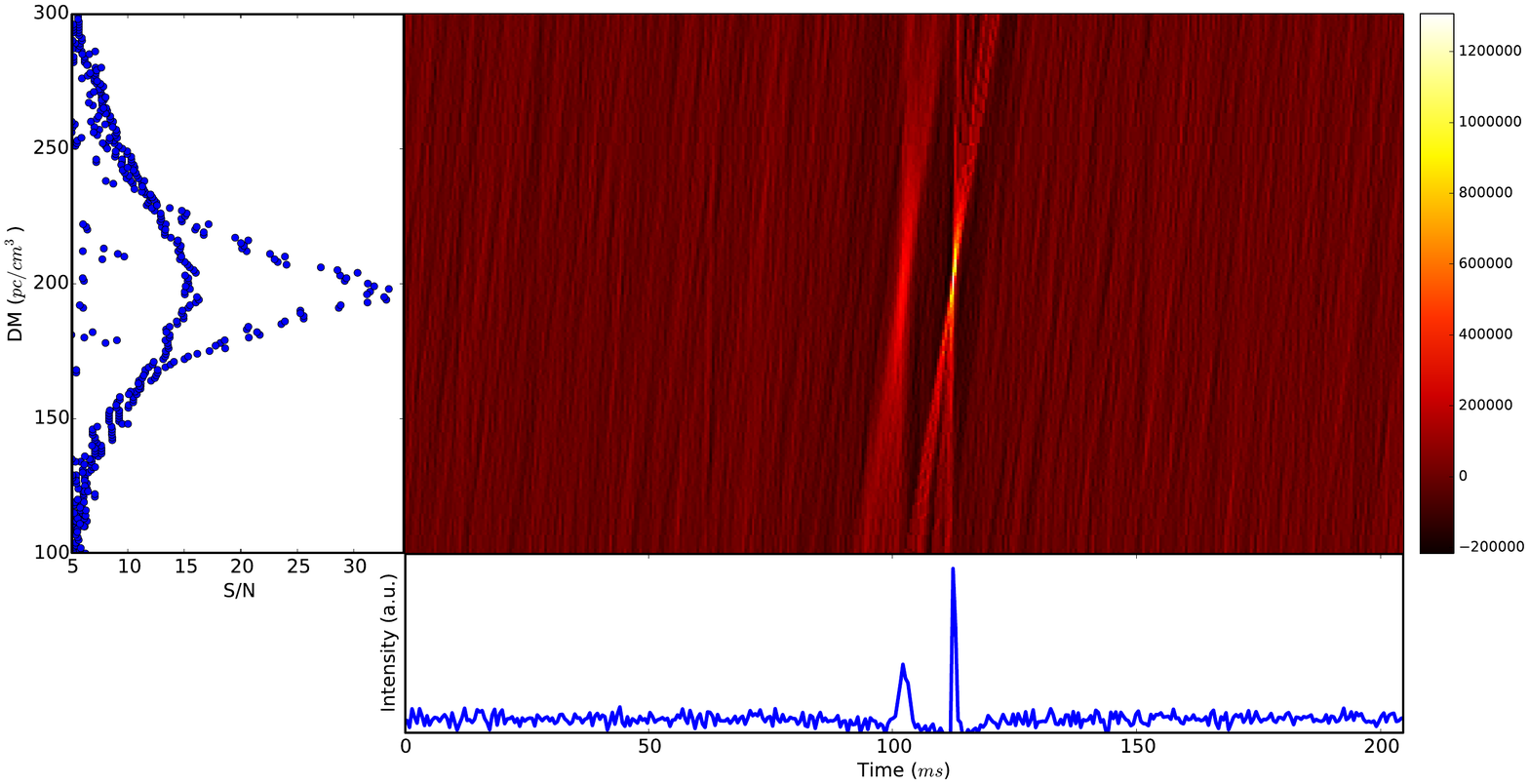} 
\caption{Above: simplified flowchart of searching for millisecond-duration pulses in e-EVN data 
in the LOCATe project. Below: bright, double pulse detected in RRAT~J1819--1458, allowing 
image-plane localization using the VLBI data.}
\label{fig6}
\end{figure}

\section{Localizing Fast Radio Bursts}

The first FRB was found in Parkes data and was reported in 2007 \cite{10}. The field progressed slowly in the 
beginning, but following the announcement of 4 Parkes FRB in 2013  \cite{44} many other instruments initiated 
millisecond pulse searches\footnote{Dan Thornton presented these results at the 2013 Lorentz Center 
workshop mentioned above, and the potential of possible EVN localization were discussed there.}. Till now
the only evidence for the extragalactic origin of these bursts is their large dispersion measure, well in excess
of the NE2001 model \cite{45} predicted Galactic value in their line of sight. But if truly distant, they could serve as a 
cosmological probe since the distribution of their DMs would tell us --~among others~--  the baryonic content of 
the Universe. The only way to prove this is direct localization using interferometric methods, and measuring the 
redshift for the optical counterpart. Recently Keane et al. reported near-real time ATCA follow-up of FRB\,150418, 
that showed a transient radio source in a galaxy with redshift $z=0.49$ \cite{46}. It has been proposed by others 
that the claimed counterpart is a scintillating AGN; indeed e-EVN data show support for the presence of an AGN 
in the galaxy (but the strong variability was not observed) \cite{47}. 

The most direct way to identify FRB counterparts is to detect these few-millisecond events in interferometric observations,
and image them. A commensal search for FRBs started at the VLBA (V-FASTR Experiment), but after four years only
upper limits on the FRB rate could be derived \cite{48}. The EVN has more sensitive dishes, but the field of view of the
larger antennas are proportionally smaller as well, thus the advantages of a similar commensal search are not straightforward.
There are however preparations within the (unfunded) LOCATe project at JIVE to establish an FRB detection pipeline. 
The idea is to search for single, dispersed pulses in autocorrelation data of the telescopes using standard pulsar software 
tools.  The data have to be RFI filtered first -- the old MkIV format EVN data caused a lot of problems initially because the 
regular data headers in the data streams were misinterpreted as RFI by the various pulsar tools. The automatic gain control
(AGC) system and the 80~Hz calibrations signal (first started at Effelsberg) are also not an advantage for single-pulse search.
The most disturbing external broad-band RFI is mitigated by a technique called zero-DM subtraction \cite{49}, also used for 
pulsar searches at Parkes. A schematic representation of the single-pulse search process is shown in Fig.~\ref{fig6}, 
along with a bright double-pulse found in our test source RRAT~J1819-1458. When a pulse is found, that tiny bit of data is 
dedispersed and recorrelated with a very fine time and frequency resolution. The following image-plane 
localization is a very simple process: one should take the calibration of the original (phase-referencing) dataset, apply
the calibration tables to the recorrelated data, and form a (series of) image(s). Since the $uv$-coverage is very poor, and the 
visibility errors in ms-pulse $uv$-data are poorly understood, one must be careful interpreting the images (Huang et al. in prep.). 

To apply this technique to FRBs has just become possible with the Arecibo discovery of repeated bursts from FRB\,121102 \cite{50}.
Following the initial VLA localization \cite{51}, the EVN provided a position to the burst source at the 10-mas level \cite{52}, showing
that it is co-located with a weak, persistent radio source in a galaxy  with a redshift of 0.1927 \cite{53}. These observations provided
the first direct evidence for the cosmological origin of a fast radio burst.

\vspace{0.25 cm}
{\small 
{\it Acknowledgements} 
The work on millisecond pulse detection with the EVN was done by the LOCATe group led by the author at JIVE; 
the most significant contributions were made by Aard Keimpema (scientific programmer), Zhigang Wen and Yuping Huang 
(summer students), and Benito Marcote (support scientist). A number of the results described in this review were reported 
and discussed during the \lq\lq Locating Astrophysical Transients'' workshop in 2013 in Leiden, which was instrumental 
for defining the future directions in transient research with the e-EVN. The organizers are grateful to the Lorentz Center 
for supporting that event. The introduction to extragalactic transients and the paragraph about isolated black holes are 
based on the (unpublished) report of this workshop to the NEXPReS e-VLBI Science Advisory Committee (eVSAG).}




\begin{thebibliography}{50}
\bibitem{1}
{\it Szomoru~A.} EXPReS and the e-EVN //
Proceedings of Science~-- 2008.~-- PoS(IX EVN Symposium)040
\bibitem{2}
{\it Reynolds~C., Paragi~Z., Garrett~M.~A.} Pipeline Processing of VLBI Data //
URSI General Assembly~-- 2002.~-- arXiv:astro-ph/0205118
\bibitem{3}
{\it Keimpema~A., et al.}  
The SFXC software correlator for very long baseline interferometry: algorithms and implementation //
Experimental Astronomy~-- 2015~-- Vol.~39(2)~-- P.~259--279
\bibitem{4}
{\it Paragi~Z., Kouveliotou~C., Garrett~M~A., et al.}
e-VLBI detection of SN2007gr //
ATel~-- 2007.~-- No.1215
\bibitem{5}
{\it Rushton~A., Spencer~R.~E., Strong~M., et al.}
First e-VLBI observations of GRS1915+105 //
Mon. Not. R. Astron. Soc.~-- 2007.~-- Vol.~374~-- L47
\bibitem{6}
{\it Tudose~V., Fender~R.~P., Garrett~M.~A., et al.}
First e-VLBI observations of Cygnus X-3 //
Mon. Not. R. Astron. Soc.~-- 2007.~-- Vol.~375~-- L11
\bibitem{7}
{\it Kulkarni~S., Kasliwal~M.}
Transients in the Local Universe //
White Paper for PTF \& LSST~-- 2009~-- arXiv:0903.0218
\bibitem{8}
{\it Paragi~Z., Godfrey~L., Reynolds~C, et al.}
Very Long Baseline Interferometry with the SKA //
Proceedings of Science~-- 2015~-- PoS (AASKA14)143
\bibitem{9}
{\it Pietka~M., Fender~R.~P., Keane~E.~F.}
The variability time-scales and brightness temperatures of radio flares from stars to supermassive black holes //
Mon. Not. R. Astron. Soc.~-- 2015.~-- Vol.~446~-- P.~3687
\bibitem{10}
{\it Lorimer~D.~R., Bailes~M., McLaughlin~M.~A., et al.}
A Bright Millisecond Radio Burst of Extragalactic Origin //
Science~-- 2007~-- Vol.~318~-- P.777
\bibitem{11}
{\it Natarayan~I., et al.}
Resolving the blazar CGRaBS J0809+5341 in the presence of telescope systematics //
 arXiv:1610:03773~-- 2016
\bibitem{12}
{\it Yang~J., Paragi~Z., van der Horst~A., et al.}
No apparent superluminal motion in the first-known jetted tidal disruption event Swift J1644+5734 //
Mon. Not. R. Astron. Soc.~-- 2016.~-- Vol.~462~-- L66
\bibitem{13}
{\it Marcaide~J.~M., et al.}
Strongly decelerated expansion of SN 1979C //
Astron. Astrophys.~-- 2002.~-- Vol.~384~-- P.~408--413
\bibitem{14}
{\it P\'erez-Torres~M.~A., et al.}
A distorted radio shell in the young supernova SN 1986J //
Mon. Not. R. Astron. Soc.~-- 2002.~-- Vol.~335~-- L23
\bibitem{15}
{\it Marcaide~J.~M., Mart\'{\i}-Vidal~I., Alberdi~A., et al.}
A decade of SN 1993J: discovery of radio wavelength effects in the expansion rate //
Astron. Astrophys.~-- 2009.~-- Vol.~505(3)~-- P.~927--945
\bibitem{16}
{\it Mart\'{\i}-Vidal~I., Marcaide~J.~M., Alberdi~A., et al.}
Radio emission of SN1993J: the complete picture. II. Simultaneous fit of expansion and radio light curve //
Astron. Astrophys.~-- 2011.~-- Vol.~526~-- A143
\bibitem{17}
{\it Parra~R., et al.}
The Radio Spectra of the Compact Sources in Arp 220: A Mixed Population of Supernovae and Supernova Remnants //
Astrophys.~J.~-- 2007~-- Vol.~659(1)~-- P.~314
\bibitem{18}
{\it P\'erez-Torres~M.~A., et al.}
An extremely prolific supernova factory in the buried nucleus of the starburst galaxy IC 694 //
Astron. Astrophys.~-- 2009~-- Vol.~507(1)~-- L17
\bibitem{19}
{\it Fenech~D., et al.}
Wide-field Global VLBI and MERLIN combined monitoring of supernova remnants in M82 //
Mon. Not. R. Astron. Soc.~-- 2010.~-- Vol.~408~-- P.~607
\bibitem{20}
{\it Muxlow~T.~W.~B., et al. }
Discovery of an unusual new radio source in the star-forming galaxy M82: 
faint supernova, supermassive black hole or an extragalactic microquasar? //
Mon. Not. R. Astron. Soc.~-- 2010.~-- Vol.~404~-- L109
\bibitem{21}
{\it Mart\'{\i}-Vidal~I., et al.}
23 GHz VLBI observations of SN 2008ax //
Astron. Astrophys.~-- 2009.~-- Vol.~499(3)~-- P.~649--652
\bibitem{22}
{\it Paragi~Z., et al.}
A mildly relativistic radio jet from the otherwise normal type Ic supernova 2007gr //
Nature~-- 2010.~-- Vol.~463~-- P.~516
\bibitem{23}
{\it van~der~Horst~A.~J., Kamble~A., Paragi~Z., et al.}
Detailed Radio View on Two Stellar Explosions and Their Host Galaxy: XRF 080109/SN 2008D and SN 2007uy in NGC 2770 //
Astrophys.~J.~-- 2011.~-- Vol.~726~-- P.~99
\bibitem{24}
{\it Soderberg~A., et al.}
A relativistic type Ibc supernova without a detected $\gamma$-ray burst //
Nature~-- 2010.~-- Vol.~463~-- P.~513--515
\bibitem{25}
{\it Pihlstr\"om~Y.~M., et al.}
Stirring the Embers: High-Sensitivity VLBI Observations of GRB~030329 //
Astrophys.~J.~-- 2007~-- Vol.~664(1), P.~411
\bibitem{26}
{\it Nappo~F., et al.}
The 999th Swift gamma-ray burst: Some like it thermal //
arXiv:1604.08204
\bibitem{27}
{\it Komossa~S.}
Tidal disruption of stars by supermassive black holes: Status of observations //
J.~High~Energy~Astrophys.~-- 2015~-- Vol.~ 7~-- P.~148
\bibitem{28}
{\it Donnarumma~I., Rossi~E.~M.}
Radio-X-Ray Synergy to Discover and Study Jetted Tidal Disruption Events //
Astrophys.~J.~-- 2015.~-- Vol.~803~-- P.~36
\bibitem{29}
{\it Romero-Ca\~nizales~C., et al.}
The TDE ASASSN-14li and Its Host Resolved at Parsec Scales with the EVN //
Astrophys.~J.~-- 2016.~-- Vol.~832~-- L10
\bibitem{30}
{\it Paragi~Z., Vermeulen~R.~C., Spencer~R.~E.}
SS433, microquasars, and other transients //
Proceedings of Science~-- 2012.~-- PoS(RTS2012)028
\bibitem{31}
{\it Vermeulen~R.~C., Schilizzi~R.~T., Spencer~R.~E., et al.}
A series of VLBI images of SS433 during the outbursts in May/June 1987 //
Astron.~Astrophys.~-- 1993~-- Vol.~270~-- P.~177
\bibitem{32}
{\it Rushton~A., et al.}
A weak compact jet in a soft state of Cygnus X-1 //
Mon. Not. R. Astron. Soc.~-- 2012.~-- Vol.~419~-- P.~3194
\bibitem{33}
{\it Paragi~Z., van~der~Horst~A.~J., Belloni~T., et al. }
VLBI observations of the shortest orbital period black hole binary, MAXI J1659--152 //
Mon. Not. R. Astron. Soc.~-- 2012.~-- Vol.~432~-- P.~1319
\bibitem{34}
{\it Merloni~A., Heinz~S., di~Matteo~T.}
A Fundamental Plane of black hole activity //
Mon. Not. R. Astron. Soc.~-- 2003.~-- Vol.~345~-- P.~1057
\bibitem{35}
{\it Spencer~R.~E.,  et al.}
Radio and X-ray observations of jet ejection in Cygnus X-2 //
Mon. Not. R. Astron. Soc.~-- 2013.~-- Vol.~435~-- L48
\bibitem{36}
{\it Deller~A.~T., et al.}
Radio Imaging Observations of PSR J1023+0038 in an LMXB State //
Astrophys.~J.~-- 2015.~-- Vol.~809(1)~-- article id.~13
\bibitem{37}
{\it Fender~R.~P., Maccarone~T.~J., Heywood~I.}
The closest black holes //
Mon. Not. R. Astron. Soc.~-- 2013.~-- Vol.~430~-- P.~1538
\bibitem{38}
{\it Porcas~R.~W., Davis~R.~J., Graham~D.~A.}
VLBI Observations of RS Ophiuchi //
\lq\lq RS Ophiuchi (1985) and the Recurrent Nova Phenomenon''. Utrecht: VNU Science Press~-- 1987~-- P.~203
\bibitem{39}
{\it Taylor~A.~R., et al.}
VLBI observations of RS OPH -- A recurrent nova with non-spherical ejection //
Mon. Not. R. Astron. Soc.~-- 1989.~-- Vol.~237~-- P.~81
\bibitem{40}
{\it O'Brien~T.~J., Bode~M.~F., Porcas~R.~W., et al.}
An asymmetric shock wave in the 2006 outburst of the recurrent nova RS Ophiuchi //
Nature~-- 2006.~-- Vol.~442(7100)~-- P.~279
\bibitem{41}
{\it Giroletti~M. et al.} 
e-EVN observations of the first gamma-ray nova V407 Cyg //
Proceedings of Science~-- 2012.~-- PoS(11th EVN Symposium)047
\bibitem{42}
{\it Chomiuk~L., et al.} 
Binary orbits as the driver of $\gamma$-ray emission and mass ejection in classical novae //
Nature~-- 2014~-- Vol.~514(7522)~-- P.~339
\bibitem{43}
{\it Miller-Jones~J.~C.~A., Sivakoff~G.~R., Knigge~C., et al.}
An Accurate Geometric Distance to the Compact Binary SS Cygni Vindicates Accretion Disc Theory //
Science~-- 2013.~-- Vol.~340(6135)~-- P.~950
\bibitem{44}
{\it Thornton~D. et al. }
A Population of Fast Radio Bursts at Cosmological Distances //
Science~-- 2013.~-- Vol.~341(6141)~-- P.~53
\bibitem{45}
{\it Cordes~J.~M., Lazio~T.~J.~W.}
NE2001. II. //
arXiv:astro-ph/0301598
\bibitem{46}
{\it Keane~E.~F., Johnston~S., Bhandari~S., et al. }
The host galaxy of a fast radio burst //
Nature~-- 2016~-- Vol.~530(7591)~-- P.~453
\bibitem{47}
{\it Giroletti~M., Marcote~B., Garrett~M.~A., et al.}
FRB 150418: clues to its nature from European VLBI Network and e-MERLIN observations //
Astron.~Astrophys.~-- 2016~-- Vol.~593~-- L16
\bibitem{48}
{\it Burke-Spolaor~S., Trott~C.~M., Brisken~W.~F., et al.}
Limits on Fast Radio Bursts from Four Years of the V-FASTR Experiment //
Astrophys.~J.~-- 2016.~-- Vol.~826(2)~-- article id.~223
\bibitem{49}
{\it Eatough~R.~P., Keane~E.~F., Lyne~A.~G.}
An interference removal technique for radio pulsar searches //
Mon. Not. R. Astron. Soc.~-- 2009.~-- Vol.~395~-- P.~410
\bibitem{50}
{\it Spitler~L.~G., Scholz~P., Hessels~J.~W.~T., et al.}
A repeating fast radio burst //
Nature~-- 2016~-- Vol.~531(7593)~-- P.~202
\bibitem{51}
{\it Chatterjee~S., Law~C.~J., Wharton~R.~S., et al.}
The direct localization of a fast radio burst and its host //
Nature~-- 2017~-- Vol.~541(7635)~-- P.~58
\bibitem{52}
{\it Marcote~B., Paragi~Z., Hessels, J.~W.~T., et al.}
 The Repeating Fast Radio Burst FRB 121102 as Seen on Milliarcsecond Angular Scales //
 Astrophys.~J.~-- 2017.~-- Vol.~834(2)~-- article id.~L8
\bibitem{53}
{\it Tendulkar~S.~P., Bassa~C.~G., Cordes~J.~M., et al.}
The Host Galaxy and Redshift of the Repeating Fast Radio Burst FRB 121102 //
Astrophys.~J.~-- 2017.~-- Vol.~834(2)~-- article id.~L7

\thispagestyle{empty}

\end{thebibliography}
\end{document}